\documentclass[twocolumn,showpacs]{revtex4}
\def\address{\affiliation}
\usepackage[dvips]{graphicx}
\begin{document}

\title{
Large In-plane Anisotropy on Resistivity and Thermopower in The Misfit Layered Oxide Bi$_{2-x}$Pb$_x$Sr$_2$Co$_2$O$_y$}

\author{
Takenori {\sc Fujii}$^{1,2}$\footnote{E-mail address:fujii@htsc.phys.waseda.ac.jp}, Ichiro {\sc Terasaki}$^{1,2}$, Takao {\sc Watanabe}$^3$ and Azusa {\sc Matsuda}$^4$.
}

\address{
$^1$Department of Applied Physics, 
Waseda University, Tokyo 169-8555, Japan.\\
$^2$Precursory Research for Embryonic Science and Technology, Japan Science and Technology Corporation, Kawaguchi 332-0012, JAPAN.\\
$^3$NTT Photonics Laboratories, Kanagawa 243-0198, JAPAN.\\
$^4$NTT Basic Research Laboratories, Kanagawa 243-0198, JAPAN.\\}%

\begin{abstract}
We investigated the in-plane anisotropy on the resistivity and thermopower of Bi$_{2-x}$Pb$_x$Sr$_2$Co$_2$O$_y$ single crystals, which have a misfit structure between the hexagonal CoO$_2$ layer and the rock salt Bi$_2$Sr$_2$O$_4$ layer. The resistivity and thermopower show significantly large anisotropy, which exceeds two at maximum. This anisotropy would come from the anisotropic pseudogap formation enhanced by the misfit structure. The thermopower changes with Pb doping to take a maximum at $x$=0.4. The misfit structure improves the thermoelectric properties through chemical pressure. The power factor is as large as 9 $\mu$W/cmK$^2$ at 100 K for $x$=0.6, which is the highest value for thermoelectric oxides at 100 K.
\end{abstract}

\date{\today}

\maketitle

The strongly correlated electron systems in oxide often show unusual physical properties, such as high-$T_c$ superconductivity in cuprates and colossal magneto-resistance in manganites. Recently, layered cobalt oxides NaCo$_2$O$_4$~\cite{terra1}, Ca-Co-O~\cite{CaCo1,CaCo2}, Bi-Sr-Co-O~\cite{BiCo,ito}, and Tl-Sr-Co-O~\cite{TlCo} were found to show a large thermoelectric power, and exibited a potential advantage for thermoelectric material. We have previously pointed out that the high thermoelectric performance of the layered cobalt oxides cannot be explained by a conventional band picture, and proposed that the strong electron-electron correlation plays an important role in the large thermoelectric power~\cite{terra2}. 

The layered cobalt oxides with large thermoelectric power consist of the alternating stack of a conducting CoO$_2$ layer and an insulating blocking layer. The Bi-Sr-Co-O system was first thought as having a structure isomorphic to that of the superconducting compound Bi$_2$Sr$_2$CaCu$_2$O$_{8+\delta}$ (Bi-2212). However, it turned out that the square Bi$_2$Sr$_2$O$_4$ lattice lies on the triangular CoO$_2$ lattice with a misfit along the $b$ axis~\cite{raveau}. Similar crystal structure was reported in ($RX$)$_x$$MX_2$ ($R$ = La, Ce, Sm, Gd, Yb; $M$= Ti, V, Nb; $X$ = S, Se), where the square $RX$ layer was intercalated between the trianglar $MX_2$ layers~\cite{Wiegers1}. These intercalated units weakly couple with conductive $MX_2$ layers via the van der Waals force. In the case of the Bi-Sr-Co-O system, the coupling between the Bi$_2$Sr$_2$O$_4$ layer and the CoO$_2$ layer is ionic and much stronger than these materials. Therefore, due to the subcells with different symmetries, the symmetry of the electronic states would be altered to induce in-plane anisotropy in physical properties. Moreover the misfit structure would induce chemical pressure along the $b$-axis direction. We further note that another structural feature in the Bi-Sr-Co-O system is a strongly modulated BiO layer, as is similarly seen in Bi-2212~\cite{matsui1} along the $b$ axis, which disappears with Pb doping~\cite{matsui2}. Thus, Bi-Sr-Co-O system is very complicated with the misfit and modulation structures. It is interesting how the electric properties are affected by the misfit and modulation structures. Here, we report on the in-plane anisotropy on the resistivity and thermopower of the Bi-Sr-Co-O system.

Single crystals of (Bi,Pb)-Sr-Co-O were grown by a traveling solvent floating zone (TSFZ) method. Starting composition was Bi$_{2-x}$Pb$_x$Sr$_2$Co$_2$O$_y$ ($x$=0, 0.4, 0.6, 0.8). The actual composition was analyzed through inductively coupled plasma-atomic emission spectroscopy (ICP) and energy dispersive X-ray analysis (EDX). Structural analysis was performed using four-circle X-ray diffractmeter (Cu K$_{\alpha}$ X-ray source) and transmission electron microscope (TEM). The resistivity was measured using a standard four probe method. The thermopower was measured using a steady-state technique.

The chemical compositions of the samples used for this study are listed against the nominal composition in Table 1(a). Hereafter, we will refer the samples as the nominal composition of Pb ($x$=0, 0.4, 0.6, and 0.8). The EDX and ICP data are in excellent agreement with each other, where (Bi+Pb) : Sr : Co $\sim$ 2.1 : 2.1 : 2.0. We can see that the actual composition of Pb is roughly equal to the nominal composition except for $x$=0.8. $x$=0.6 and $x$=0.8 exhibit no difference in the actual chemical composition, the lattice constant, and the transport properties, which indicates that the solubility limit of Pb is considered to be about $x$=0.6. 

The lattice constants and the angle $\beta$ are listed in Table 1(b). With increasing Pb concentration, the $c$-axis length expands continuously from 29.84 \AA of $x$=0 to 30.04 \AA of $x$=0.6, whereas the $b$-axis length of the rock-salt layer $b_{RS}$ discontinuously shrinks from 5.38 \AA of $x$=0 to 5.22 \AA of $x$=0.4 with the $a$-axis length of the rock-salt layer $a_{RS}$ nearly unchanged (4.9 \AA). This is in good agreement with the powder XRD measurement by Yamamoto $et$ $al$~\cite{yamamoto}.
\begin{table}
\caption{\label{tab:table1}(a) Chemical composition estimated from energy dispersive X-ray (EDX) analysis and inductively coupled plasma-atomic emission spectroscopy (ICP). (b) Lattice parameter determined from four-circle X-ray diffrectmater. $b_{RS}$ and $b_H$ are the $b$-axis length of the rock-salt and the hexagonal subcells, respectively. $b_H$ is determined by the electron diffraction patterns.}
\begin{ruledtabular}
\begin{tabular}{@{\hspace{\tabcolsep}%
				\extracolsep{\fill}}ccccccc}
\multicolumn{3}{l}{(a) Composition}\\ 
\\ 
Sample & & Bi & Pb & Sr & Co \\ 
\hline 
$x$=0 &EDX  & 2.21   &   0 & 2.20 & 2.0 \\ 
 & ICP &  2.1  & 0& 2.1&2.0 \\ %
$x$=0.4 &  & 1.61  &   0.40&2.09&2.0  \\ 
 &  &1.82   &   0.42&2.18&2.0 \\ 
$x$=0.6 & & 1.58    & 0.57&2.13&2.0 \\ 
 &  & 1.60   &  0.60&2.10&2.0  \\ 
$x$=0.8 & & 1.60 &0.55 & 2.15 &2.0  \\ 
&&-&-&-&-\\ 
\end{tabular}
\end{ruledtabular}
\end{table}
\begin{table}
\begin{ruledtabular}
\begin{tabular}{@{\hspace{\tabcolsep}%
				\extracolsep{\fill}}ccccccc}
\multicolumn{3}{l}{(b) Lattice parameter}\\ 
\\
Sample & $a_{RS}$ (\AA) & $b_{RS}$ (\AA) &$c$ (\AA) & $b_H$ (\AA) & $\beta$ (deg.) \\
\hline
$x$=0 &  4.937 & 5.405 & 29.875 & 2.8 & 93.554\\
$x$=0.4&   4.914  &  5.221&29.974&2.8&92.315  \\ 
$x$=0.6&  4.904&5.206&30.041&-&92.657   \\
\end{tabular}
\end{ruledtabular}
\end{table}

Figures 1(a) and (b) show the TEM diffraction patterns of $x$=0 and 0.4, respectively, which clearly show a rock-salt diffraction pattern from the Bi$_2$Sr$_2$O$_4$ layer and a hexagonal diffraction pattern from the CoO$_2$ layer. The $a$- and $b$-axis length of the hexagonal CoO$_2$ layer ($a_{H}$, $b_{H}$) is estimated to be about 2.8 \AA and the angle between them is about 60$^{\circ}$, which is independent of $x$ within the resolution limit of the TEM image. From the above structural analysis, the prepared samples show the misfit structure along the $b$ axis (13 $b_{H}$ $\sim$ 7 $b_{RS}$ $\sim$ 36 \AA), while the $a$-axis length of the rock-salt layer matches with that of the hexagonal layer ($a_{RS}$ $\sim$ $\sqrt{3}$ $a_{H}$ $\sim$ 4.9 \AA). The TEM diffraction pattern of $x$=0 shows satellite reflection due to the modulation structure along the oblique direction (tilted about 45$^\circ$) from $a^*_{RS}$ or $b^*_{RS}$. In contrast, there is no satellite reflection in $x$=0.4, indicating that the modulation structure disappears in this compound. This modulation structure was also confirmed with the Laue transmission photographs (not shown here).

\begin{figure}
 \includegraphics[width=6cm,clip]{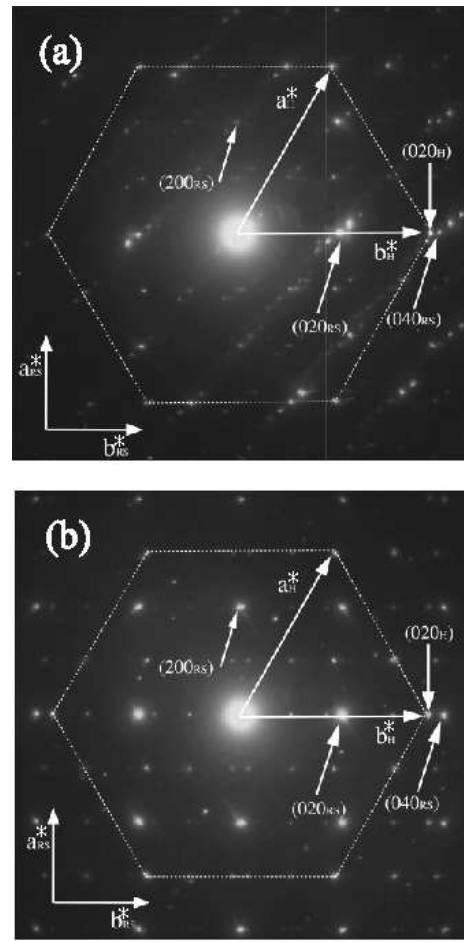}
 \caption{
 Electron diffraction patterns of (a) $x$=0 and (b) $x$=0.4.
 }
\end{figure}

Figure 2 shows the temperature dependence of the $a$- and $b$-axis resistivities (We refer the $a$- and $b$-axis directions as the $a_{RS}$ and $b_{RS}$ directions). All the samples show a metallic behavior at room temperature. With Pb doping, the magnitude of the resistivity continuously decreases, suggesting that carriers are doped through the substitution of divalent Pb$^{2+}$ for trivalent Bi$^{3+}$~\cite{yamamoto,ito}. Note that the magnitude of resistivity for all samples is unusually large for a metal. The large $\rho$ comes from a short in-plane mean free path ($\sim$ 3 \AA at 300 K) which is comprable to the Co-Co bond length, and suggests the incherent nature of the metallic behavior (These materials are called bad metal)~\cite{emery}.

The in-plane anisotropy of the resistivity ($\rho_b$/$\rho_a$) is shown in the inset of Fig. 2. Contrary to the relatively small $\rho_b$/$\rho_a$ for $x$=0, $\rho_b$/$\rho_a$ of $x$=0.4 increases strongly below 50 K. This comes from the different temperature dependence; $\rho_b$ shows an upturn below 50 K, whereas $\rho_a$ remains metallic down to 4.2 K. Pb substitution for Bi strongly suppresses the low-temperature upturn and $x$=0.6 is again less anisotropic. (Since $\rho_a$ is sensitive to the current direction, a small upturn of $\rho_a$ of $x$=0.6 is possibly due to the misalignment of the contacts.) Since the triangular CoO$_2$ layer itself shows no in-plane anisotropy, the observed anisotropy is anomalously large from the viewpoint of group theories. This is considered to come from the rock-salt Bi$_2$Sr$_2$O$_4$ layers of different symmetry. Actually, Bi-2212, which consists of the square CuO$_2$ and the rock-salt Bi$_2$Sr$_2$O$_4$ layers, shows much less in-plane anisotropy in the resistivity~\cite{fujii}. We have previously proposed the pseudogap formation in Bi-Sr-Co-O at low temperatures~\cite{ito}, and have attributed the upturn of the low temperature resistivity to the decrease in the density of states. Accordingly, the in-plane anisotropy suggests that the pseudogap is anisotropic (nearly zero along the $a$ axis).

\begin{figure}
 \includegraphics[width=7cm,clip]{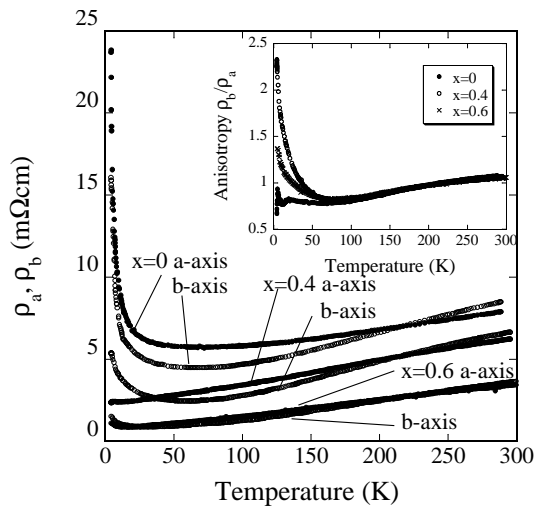}
 \caption{
 Temperature dependence of the $a$- and $b$-axis resistivities. Inset: In-plane ansotropy of the resistivity ($\rho_b$/$\rho_a$).
 }
\end{figure}

Figures 3(a), (b) and (c) show the temperature dependence of the $a$ and $b$ axis thermopower for $x$=0, 0.4 and 0.6 respectively. Pb substitution for Bi enhances the thermopower in both directions from $x$=0 to 0.4. For example, the magnitude of the $a$-axis thermopower increases from 140 to 160 $\mu$V/K at room temperature in going from $x$=0 to 0.4. However, it decreases down to 140 $\mu$V/K for $x$=0.6. We have claimed that the large thermopower in Co oxides is attributed to the strong electron-electron correlation~\cite{ito}. The increase of the thermopower from $x$=0 to 0.4 is considered to be due to the discontinuous shrink of the $b_{RS}$ axis, which causes a chemical pressure along the $b$ axis. A similar phenomenon is seen in the Ce-based compounds~\cite{terra2}, where the thermopower increases with increasing pressure~\cite{jaccard}. The decrease of the thermopower from $x$=0.4 to 0.6 is possibly due to the increase of the carrier density, because the $b$-axis length is nearly unchanged.

Next, we consider the anisotropy in the thermopower. In the inset of Fig. 3(b), we show the in-plane anisotropy of the thermopower ($S_b$/$S_a$). All the samples exhibit a large anisotropy which is characterized by a larger $b$-axis thermopower below 100 K. At around 50K, the $b$-axis thermopower of $x$=0.4 is about two times larger than the $a$-axis thermopower. Note that the temperature dependence of $S_b/S_a$ roughly corresponds to that of $\rho_b/\rho_a$. The magnitude of $S_b/S_a$ also agrees with that of $\rho_b/\rho_a$, where $x$=0.4 shows the largest value of two. These results strongly suggest that $S_b/S_a$ and $\rho_b/\rho_a$ are predominantly determined by the density of states, and that the scattering time is essentially temperature-independent, which is consistent with the bad-metal picture. In this context, $S_b/S_a$ comes from the anisotropic density of states induced by the anisotropic pseudogap. One may notice that the anisotropy of the resistivity is smaller than that of the thermopower for $x$=0. We attribute this to the modulation structure, which works as anistropic scattering center for Bi-2212~\cite{fujii}. 

\begin{figure}
 \includegraphics[width=7cm,clip]{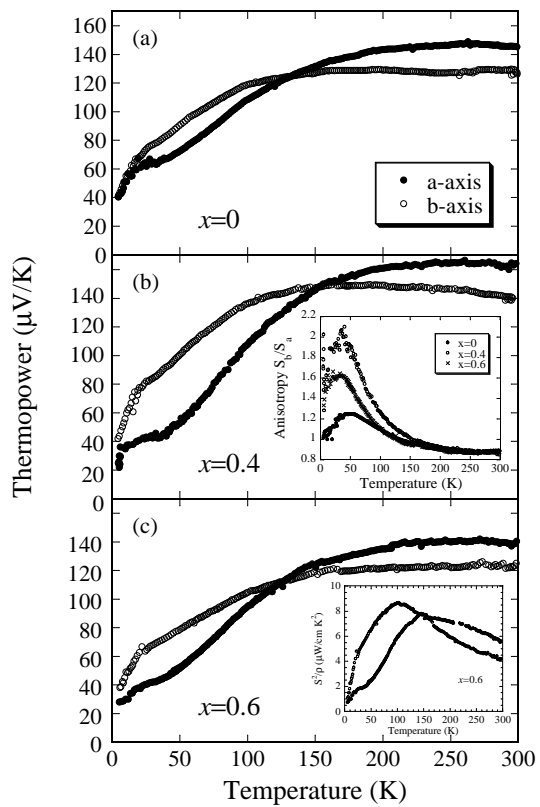}
 \caption{
 Temperature dependence of the $a$- and $b$-axis thermopower for (a) $x$=0, (b) $x$=0.4, and (c) $x$=0.6. Inset of (b): In-plane anisotropy of the thremopower ($S_b$/$S_a$). Inset of (c): The power factor ($S^2/\rho$) of $x$=0.6 along the $a$- and $b$- directions.
 }
\end{figure}

The power factor ($S^2/\rho$) is a measure of the thermoelectric properties, which means that large themopower and low resistivity are required to thermoelectric materials. The power factor along the $b$ axis for $x$=0.6 is more than two times as large as that along $a$ axis at around 50 K (Inset of Fig. 3(c)), which indicates that the thermoelectric properties can be improved (doubled in the present case) with controlling the misfit structure. The magnitude reaches as large as 9 $\mu$W/cmK$^2$ that is highest for thermoelectric oxides at 100 K.

Finally, let us discuss the origin of the anisotropic pseudogap. We have previously proposed that a spin-density-wave (SDW)-like state is a possible origin for the pseudogap in NaCo$_2$O$_4$~\cite{terra3}. SDW is closely related to the topology of the Fermi surface and its nesting, and thus the pseudogap is favorable to open along a nesting vector. Assuming that the hexagon-like Fermi surface calculated for NaCo$_2$O$_4$~\cite{shin} is also applicable for the Bi-Sr-Co-O system, we expect that SDW is formed along the six-fold $\Gamma$-K direction. The misfit structure lowers the crystal symmetry of the CoO$_2$ layer to suppress the nesting along the $a$-axis direction. Considering that disorder induces an SDW-like transition in NaCo$_2$O$_4$, we think that the misfit structure is likely to enhance the SDW instability in the Bi-Sr-Co-O system. 

Another scenario is based on the fact that the in-plane anisotropy of the Bi-Sr-Co-O system is as large as that of YBa$_2$Cu$_3$O$_y$ for the heavily oxygenated sample of $y$=7 and the oxygen deficient sample of $y$=6.35~\cite{ando}. There are additional conductive CuO chains in $y$=7, which are responsible for the large anisotropy. With decreasing oxygen content, the CuO chains are progressively destroyed, and the orthorhombicity almost vanishes in $y$=6.35. The large anisotropy in $y$=6.35 is attributed to a self-organization of the two-dimensional electrons, such as a charge stripe. In this sense, there is a possibility that the in-plane anisotropy in the Bi-Sr-Co-O system comes from a charge stripe.

In conclusion, we grew single crystals of Bi$_{2-x}$Pb$_x$Sr$_2$Co$_2$O$_y$ by a TSFZ method. From the structural analysis, we confirmed that they have a misfit structure along the $b$ axis, and that $x$=0 exhibits a modulation structure, whose direction is tilted by about 45$^\circ$ from the $a$- and $b$-axes. We found that the resistivity and thermopower shows a large anisotropy, which would come from the anisotropic pseudogap formation. Accordingly the pseudogap in the $b_{RS}$-axis direction makes the resistivity nonmetallic, and the thermopower larger at low temperatures. Aside from the pseudogap, the misfit structure enhances the thermopower through chemical pressure, and we suggest that we can improve the thermoelectric properties by controlling the misfit structure. The maximum power factor is as large as 9 $\mu$W/cmK$^2$ at 100 K for $x$=0.6 along the $b$ axis, which is the highest value at 100 K among various thermoelectric oxides.

\end{document}